\documentclass[prd, preprint, showpacs,amsmath,amssymb]{revtex4-1}
\usepackage{graphicx}
\usepackage{epsfig}
\usepackage{amssymb}
\usepackage{graphicx}
\usepackage{dcolumn}
\usepackage{bm}
\newcommand{\ba}{\begin{array}}
\newcommand{\ea}{\end{array}}
\def\br{\begin{eqnarray}}
\def\er{\end{eqnarray}}
\def\be{\begin{equation}}
\def\ee{\end{equation}}

\def\({\left(}
\def\){\right)}

\begin{document}


\title{Scalar bosons in Minimal and Ultraminimal Technicolor: Masses, trilinear couplings and widths}

\author{A.~Doff$^1$ and A.~A.~Natale$^2$}
\affiliation{$^1$Universidade Tecnol\'ogica Federal do Paran\'a - UTFPR - COMAT
Via do Conhecimento Km 01, 85503-390, Pato Branco - PR, Brazil \\
$^2$Instituto de F\'{\i}sica Te\'orica, UNESP - Univ. Estadual Paulista,
Rua Dr. Bento T. Ferraz, 271, Bloco II,
01140-070, S\~ao Paulo - SP,
Brazil }

\date{\today}

\begin{abstract}
We compute masses, trilinear self-couplings and decay widths into weak bosons of the scalar composite bosons in the case of the Minimal and Ultraminimal technicolor models. The masses, computed via the Bethe-Salpeter equation, turn out to be light and the trilinear couplings smaller than the one that would be expected when compared to a fundamental Standard Model scalar boson with the same mass. The decay widths into electroweak bosons of the Ultraminimal model scalars bosons are much smaller than the one of the Minimal model.
\end{abstract}


\maketitle

\section{Introduction}

The understanding of the gauge electroweak symmetry breaking mechanism is one of the most important
problems in particle physics at present. One of the explanations of this mechanism is based on new strong interactions usually named technicolor (TC). The early technicolor models \cite{wei} 
suffered from problems like flavor changing neutral currents (FCNC) and contribution to the electroweak corrections not
compatible with the experimental data, as can be seen in the reviews of Ref.\cite{tc}. However the TC dynamics
may be quite different from the known strong interaction theory, i.e. QCD, this fact has led to the walking
TC proposal \cite{walk}, which are theories where the incompatibility with the experimental data has been solved,
making the new strong interaction almost conformal and changing appreciably its dynamical behavior. We can obtain an almost conformal TC theory, when the fermions are in the fundamental representation, introducing a large number of TC fermions ($n_F$), leading to an almost zero $\beta$ function and flat asymptotic coupling constant. The cost of such procedure may be a large S parameter \cite{peskin} incompatible with the high precision electroweak
measurements. 

Recently a Minimal(MWT) \cite{sannino1} and an Ultraminimal(UMT) \cite{sannino2} TC models were proposed
where the presence of TC fermions in other representations than the fundamental one led to viable models without conflict
with the known value for the measured S parameter. An effective Lagrangian
analysis indicates that such models also imply in a light scalar Higgs boson \cite{sannino1,sannino2,sanninonew,sannino3}. This possibility was investigated
and confirmed by us through the use of an effective potential for composite operators \cite{doff1} and through a calculation involving the Bethe-Salpeter equation (BSE) for the scalar state \cite{doff2}. The BSE approach to compute the scalar masses is a straightforward one, and our purpose in this paper is complement the studies of dynamical symmetry breaking of Refs.\cite{doff1,doff2} in the case of the specific Minimal and Ultraminimal  TC models. Moreover in Ref.\cite{doff1,doff3} we also estimated the trilinear
scalar self-coupling which also could be measured in the case of light Higgs bosons \cite{eboli}. Our main
result is a table where we indicate the scalar masses, trilinear self-couplings and decay widths into
electroweak bosons for these models which can be confronted with the experiment in the case that a TC composite scalar boson is found at LHC.

This paper is organized as follows: In Sec. II  we specify the fermionic content  and we obtain  the fermionic Schwinger-Dyson equations, or gap equation, for the Minimal and Ultraminimal TC models. In  Section III we compute  for these models masses, trilinear self-couplings and decay widths into weak bosons of the scalar composite bosons. In Sec. IV we draw our conclusions.

\section{Fermionic Schwinger-Dyson equations for the MWT and UMT models}

The Minimal TC model is based on a $SU(2)$ gauge group with two adjoint fermions \cite{sannino1}
\begin{equation}
Q^a_L  =	\left( \ba{c}
	      		U^a\\
	      		D^a
              \ea \right)_L  \;\; \; , \;\;\; U^a_R , \;\;\; D^a_R , \;\;\; a=1, 2, 3,
\label{eq1}
\end{equation}
where $a$ is the $SU(2)$ adjoint color index and the left-handed fields correspond to three ($SU(2)_L$) weak
doublets. The Ultraminimal TC model is based on a two colors group with two fundamental Dirac flavors 
$SU(2)_L\times U(1)_Y$ charged described by \cite{sannino2}
\begin{equation}
T_L  =	\left( \ba{c}
	      		U\\
	      		D
              \ea \right)_L  \;\; \; , \;\;\; U_R , \;\;\; D_R ,
\label{eq2}
\end{equation}
and also two adjoint Weyl fermions indicated by $\lambda^f$ with $f=1,2$, where these fermions are  singlets under  $SU(2)_L\times U(1)_Y$. 

The near conformal behavior for these models can be observed looking at the zero of the two-loop $\beta (g^2)$ function, which is given by
\be
\beta (g) = -\beta_0 \frac{g^3}{(4\pi)^2} - \beta_1 \frac{g^5}{(4\pi)^4} \,\,\, ,
\label{eq01}
\ee
where
\be
\beta_0 = (4\pi)^2 b= \frac{11}{3} C_2(G) - \frac{4}{3}T(R)n_F (R) \,\,\, , 
\label{eq02}
\ee
and 
\be
\beta_1 = \left[\frac{34}{3}C_2^2(G)-\frac{20}{3}C_2(G)T(R)n_F -4C_2(R)T(R)n_F \right] ,
\label{eq03}
\ee 
where $C_2(R)I=T^a_RT^a_R$, $C_2(R)d(R)=T(R)d(G)$, $d(R)$ is the dimension of the representation $R$ and
$G$ indicates the adjoint representation. It is interesting to compare the leading term of the $\beta$ function for the different models (indicated
respectively by $b_{mi}$ and $b_{um}$, while the one of a simple walking TC theory is denoted by $b_w$). In the case of an $SU(2)$ gauge group with $8$ Dirac fermions we have $b_w = 2/16\pi^2$. In the Minimal walking model we obtain the same coefficient with only $2$ fermions ($b_w=b_{mi}$)! The main difference among these models appears when we compute the S parameter whose
perturbative expression (in the massless limit) is
\be
S=\frac{1}{6\pi}\frac{n_F}{2}d(R) \,\, .
\label{s1}
\ee
The data requires the value of the S parameter to be less than about 0.3. According to the ``naive'' perturbative
estimate of Eq.(\ref{s1}) this requirement is indeed met for MWT (and also for UMT). However, early models have fermions
 only in the fundamental representation of SU(2) or SU(3). Then one needs quite a large $n_F$ to make the theory walking, 
 and therefore the perturbative estimate of S contradicts with data.

The main difficulty in TC lies in the small knowledge that we have about the chiral symmetry breaking
pattern of such strongly interacting theories. In the models that we will discuss here most of the
information about the chiral symmetry breaking comes from the use of effective theories \cite{sannino1,sannino2} and the effective potential generated by them \cite{sannino3}. Another way to unravel the symmetry
breaking pattern in TC theories is through the effective potential for composite operators
as computed recently in Ref.\cite{doff1}. It is also possible to obtain information on the spectrum of TC theories simply looking at the gap equations and their possible solutions, this is the simplest approach and the point
of view to be followed here using some of the results of Ref.\cite{doff2}. Of course, all these
attempts involve a reasonable uncertainty typical of non-perturbative theories, but the full set of 
results may be able to corner the main characteristics of the broken TC theory, i.e. masses and couplings. 

In order to discuss masses and couplings, as performed in Ref.\cite{doff1,doff2}, we need to know the
solution of the fermionic Schwinger-Dyson equations, or gap equation, for the Minimal and Ultraminimal TC models. 
The two basic parameters that define the gap equation are: The $\beta$ function coefficients appearing in the coupling constant and the Casimir operators resulting from the fermion-gauge boson vertex.
The gap equation, for fermions in the representation $R$, can be written as
\br
\Sigma (k) &=& \frac{3C_2(R)}{16\pi^2}\left[ \frac{g^2(k^2)}{k^2}\int_0^{k^2} \right.
\frac{p^2dp^2\Sigma(p^2)}{p^2+m^2}
 +  \left. \int^\infty_{k^2} \frac{dp^2g^2(p^2)\Sigma(p^2)}{p^2+m^2} \right],
\label{eq3}
\er 
where the coupling constant $g^2(p^2)$ behaves as
\be
g^2(p^2)=\frac{g^2(m^2)}{1+bg^2(m^2)ln\left(\frac{p^2}{m^2}\right)},
\label{eq4}
\ee
and where $m\approx\Lambda_{TC}$ is the dynamical mass scale that is assumed to be equal to the TC scale. 
The factor $b$ in Eq.(\ref{eq4}), is the
one that comes out from the full behavior of the $\beta$ function in Eqs.(\ref{eq01}-\ref{eq03}) (i.e., including
all fermionic representations). Eq.(7) may not be the full equation that reflects all the possible SDE solutions. There are
possible contributions that may modify this equation. For instance, there may be strong non-perturbative effects at short distances, generated
by extended technicolor or other new interactions, that may produce effective four fermion
interactions leading to a behavior that we call ``extreme walking", which is the one that reduces the anomalous dimension of the operator ${\bar{\Psi}}\Psi$ to $1$, and which has been termed as a Nambu-Jona-Lasinio (NJL) limit. Unfortunately it is nowdays known that the gap equation cannot explain even the QCD's  chiral symmetry breaking (Ref.\cite{cornwall} discuss the importance of confinement for this symmetry breaking)! 
For example, if we consider the most recent QCD lattice simulations leading to infrared
finite gluon propagators, we certainly do not obtain chiral symmetry breaking from the gap equation at all! Taking
this fact (finite gauge boson propagators)  into account when we consider TC theories, we observe that the gauge boson mass scale erase the strength
necessary for the chiral symmetry breaking\cite{pn}, and only in the NJL limit we can obtain chiral symmetry breaking in the gap equation for fermions in the fundamental representation. It is also interesting to notice that dynamical gauge boson masses also imply the existence of a non-trivial fixed point\cite{cnp}, what may be in agreement at some extent with the expected behavior of a ``walking" theory, but all these points are missing in all TC gap equation calculations. Therefore our approach will be a phenomenological one, and, in the sequence, when we discuss any solution for the fermionic SDE we mean all possible solutions of Eq.(7) with the addition of all possible corrections due to some unknown dynamics. A quite general solution for Eq.(\ref{eq3}) is \cite{doff4,doff1}
\be 
\Sigma (p) = m\left(\frac{m^2}{p^2}\right)^{\alpha}\left[ 1 + bg^2\ln\left(\frac{p^2}{m^2}\right)\right]^{-\gamma(\alpha)} \,\, ,
\label{eq5}
\ee
where $\gamma (\alpha) = \gamma \cos{(\alpha \pi)}$ and
\[
\gamma= \frac{3c}{16\pi^2 b}  \,\,\, .
\] 
$c$ is the quadratic Casimir operator given by 
\be
c = \frac{1}{2}\left[C_{2}(R_{1}) +  C_{2}(R_{2}) - C_{2}(R_{3})\right] \,\,\, ,
\label{eq6}
\ee
where $C_{2}(R_{i})$,  are the Casimir operators for fermions in the representations  $R_{1}$ and 
$R_{2}$ that form a composite boson in the representation $R_{3}$. If $R_1=R_2=R$ and $R_3$ is
the singlet state $c$ is simply reduced to $C_2(R)$. 
The only restriction on this solution is $\gamma > 1/2$ \cite{lane}. This solution can be 
understood as one ansatz that maps all possible behavior of the gap equation as we vary $\alpha$,
since the standard operator product expansion (OPE) behavior for $\Sigma (p^2)$ is 
obtained when $\alpha \rightarrow 1$, whereas the ``extreme walking" TC solution is obtained when $\alpha \rightarrow 0$. As explained at length in Refs.\cite{doff1,doff4} most of the calculations can be performed
with the expression of Eq.(\ref{eq5}) and afterward we consider the ``extreme walking" (or NJL) limit $\alpha =0$. Note that this is
the only possible solution that is naturally able to reproduce the top quark mass\cite{doff4}.
In the Minimal TC model we have fermions in the adjoint representation and $R_1=R_2=G$ in
Eq.(\ref{eq6}), while in the Ultraminimal TC model the factor $c$, and consequently $\gamma$,
have to change in order to consider the Casimir operators of the Dirac fermions in
the fundamental representation ($c_F$) and the Weyl fermions in the adjoint representation ($c_G$), with
condensation occurring in the TC singlet channel.   

\section{The scalar bosons masses, trilinear couplings and decay widths in the MWT and UMT models }

In Ref.\cite{doff2} we obtained the scalar boson mass in the  case of an ``extreme walking" TC theory through
a calculation based on the BSE, which is given by (see Eq.(26) of Ref.\cite{doff2})
\br
M_{H}^{2(0)}  &\approx & 4 v^2\left(\frac{8 \pi^2  bg^2(m)(2\gamma -1)}{N_{TC}n_F} \right)A  ,
\label{eqf}
\er
where  
$$
A = \left( \frac{1}{4}\frac{bg^2(m)(2\gamma - 1)}{(1 + \frac{bg^2(m)(2\gamma - 1)}{2})}\right)
$$
\noindent and  $v \sim 246 GeV$ is the Standard Model vacuum expectation value (vev) and we are considering a $SU(N_{TC})$ group. This equation depends only on the electroweak group vev and on the group theoretical factors and number of fermion flavors. Note that Eq.(\ref{eqf}) indicates that the scalar masses are lowered in quasi-conformal gauge theories as a consequence of the BSE normalization condition as discussed in Ref.\cite{doff2}.

It is interesting to verify if Extended Technicolor (ETC) can change our predictions for the scalar masses. ETC does introduce some chiral symmetry breaking in the TC sector (current technifermion masses $m_{TC}\neq 0$), which may affect strongly the techni-pion spectra (through Dashen's relation
$m_{TC}<{\bar{\Psi}}_{TC}\Psi_{TC}>=m^2_\Pi F^2_\Pi$) when we take into account specific details of the complete model. However the scalar masses in dynamical symmetry breaking models are more related to the dynamical masses than to the current techni-fermion mass, i.e. 
\[
M_H \approx 2 m_{dyn}^{TC} A  \,\, .
\]
This relation of the scalar mass with the dynamical one (apart from the factor $A$) goes back to the work of Nambu and Jona-Lasinio and was shown to work in QCD by Delbourgo and Scadron\cite{scadron}. The factor $A$ takes into account the normalization of the Bethe-Salpeter equation that was neglected until recently\cite{doff2}. We can therefore assume that $ m_{dyn}^{TC}$ is the full dynamical mass self-generated by TC with the addition of a current mass generated through ETC,  i.e. the Schwinger-Dyson equation (SDE) for the dynamical mass can be written as
\be
\Sigma^\prime (k^2)= \Sigma_{TC} (k^2)+\delta\Sigma (k^2) \,\, ,
\ee
where $\delta\Sigma (k^2)$ denotes the ETC contribution to the TC fermion mass, whose SDE can be approximated by
\be
\delta\Sigma (0) \approx \frac{3c_{{}_{ETC}}\alpha_{{}_{ETC}}}{4\pi \Lambda^2_{ETC}} \int^{\Lambda^2_{ETC}} 
\frac{p^2dp^2 \,\Sigma_{TC}(p^2)}{p^2+\Sigma_{TC}^2(p^2)} \,\, .
\ee
We have cut the integral at the asymptotic limit $\Lambda^2_{ETC}$. Assuming the extreme walking behavior we can estimate the behavior of $\delta\Sigma (0)$, which will result in
\be
\delta \Sigma(0)\approx  \frac{1}{4}\frac{\alpha_{{}_{ETC}}c_{{}_{ETC}}}{\alpha_{{}_{TC}}c_{{}_{TC}}}m\approx
\frac{3}{4\pi}\alpha_{{}_{ETC}}c_{{}_{ETC}}m \,\, ,
\label{eqx}
\ee 
where $m\propto \Sigma_{TC} (p^2\rightarrow 0)$. Also assuming the most attractive channel hypothesis ($\alpha_{{}_{TC}}c_{{}_{TC}} \approx \pi/3$)  and considering that $\alpha_{{}_{ETC}}$ in general is a small number we see that Eq.(\ref{eqx}) leads to a small correction for the scalar mass. 
As one example we could refer to a model proposed by Appelquist and Schrock\cite{Appelquist}, where the largest ETC contribution comes from a third stage of symmetry breaking at one scale ($\Lambda_{{}_{ETC}} = \Lambda_3 = 10$ TeV) and with a coupling $\alpha_{{}_{ETC}} = \alpha_3 = 0.4$.
In this particular case the correction to the scalar mass is quite small. We investigated other models and in none of them the corrections are larger than $10\%$. Therefore we can neglect ETC corrections to the scalar mass when compared to the other uncertainties involved in this problem.

The scalar composite coupling to the ordinary quarks is determined through Ward identities as
discussed in Ref.\cite{soni,doff3}
\be
\imath \lambda_{Hff} \propto -\imath \frac{g_W \Sigma (k)}{2M_W} \,\,\, ,
\label{eq28}
\ee
where $g_W$ and $M_W$ are, respectively, the weak coupling constant and gauge boson mass. A formal demonstration how such coupling is found
can be seen in the papers of Ref.\cite{Cornwall} .

\begin{figure}[t]
\centering
\includegraphics[width=0.4\columnwidth]{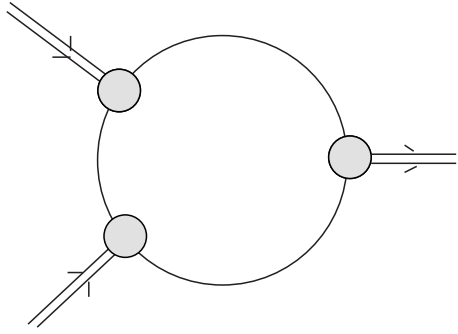}
\caption[dummy0]{Top quark (fermionic internal lines) contribution to the trilinear composite ($3H$) Higgs boson coupling. The gray blobs are proportional to the effective $ttH$ coupling. }
\label{fig:H1}
\end{figure}

With the help of Eq.(\ref{eq28}) we can compute the trilinear self-coupling which is giving by the diagram of Fig.(\ref{fig:H1}).
This diagram is dominated by the heaviest ordinary fermion (the top quark) and no contribution
comes out from the TC quarks, as discussed in Ref.\cite{doff1}, where the following coupling
was obtained 
\be
\lambda^{(0)}_{3H} \approx \frac{9g^3_{W}}{32\pi^2}\frac{m_{t}}{bg^2(4\gamma - 1)}\left(\frac{m_{t}}{M_{W}}\right)^3\left[1 - \frac{4\alpha}{bg^2(4\gamma - 2)}+ ...\right] \, ,  
\label{eq29}
\ee
where $m_t$ is the top quark mass.

The decay width into electroweak gauge bosons was calculated in the second paper of Ref.\cite{doff2}
and, in the $SU(2)$ linear sigma model approximation, is given by
\be
\Gamma_{H\rightarrow WW}\approx \frac{3m^3}{4F_\Pi^2} \,\, ,
\label{larg}
\ee
where the relation $m^3/F_\Pi^2$ can be written in terms of the electroweak group vev, the group theoretical factors and the number of fermion flavors \cite{doff1,doff2}.

All the above expressions were obtained assuming the most attractive channel hypothesis ($cg^2/4\pi \approx 1$) and
that the technifermions are in an unique representation. They cannot be applied straightforwardly for the Ultraminimal TC model, because
in this case we have two scalar composite bosons, that may appear as mixed states formed by fermions in the fundamental and adjoint representations.
For instance, the Ultraminimal gap equation has two contributions, one with a Casimir operator for fermions in the fundamental representation and another with a different Casimir  operator for fermions in the adjoint representation, while it is
the same $\beta$ function that governs the running of the coupling in the two contributions.   
It is opportune to remember that the gap equation lead to different chiral symmetry breaking scales when the fermions
are in different representations, this has been observed, for instance, in QCD with quarks in
the adjoint representation \cite{cornwall}, where the chiral transition may be slightly different from the
confinement transition, which coincides with the chiral one for fermions in the fundamental representation.
We expect that the masses and composite scalar wave functions will show a mixing but with scales quite
close to the TC scale ($\Lambda_{TC}$).
The expression for the decay width into electroweak bosons must also be modified because the
decay is proportional only to the scalar wave function formed by TC fermions in the fundamental representation,
once the technifermions in the adjoint representation are singlets under the electroweak group.
The scalar bosons will mix among themselves and quite probably with scalars formed by
technigluons. The determination of the mixing angle in a problem with two composite (or more) strongly interacting bosons is a quite difficult problem and it will not be considered here. 

We shall make use of a trick based on the behavior of the
Bethe-Salpeter scalar wave function to estimate the magnitude of masses and decay widths
of the Ultraminimal scalar bosons.
The Bethe-Salpeter wave function for the scalar composite boson in the limit that the internal momentum
$q_\mu \rightarrow 0$ can be expressed by
\be
\chi (p) = S(p)\gamma_5 \frac{\Sigma (p)}{F_{\Pi}}S(p) \,\, ,
\label{bs}
\ee
where $S(p)$ is the fermion propagator. In this limit this equation is known to have the same
solution as the fermionic self-energy, is dependent on the $\beta$ function coefficients
and the Casimir operators in the anomalous dimension, and it is possible to verify that
the $0^+$ wave functions formed by fermions in the fundamental and adjoint representation, as
happens in the Ultraminimal model, scale with the Casimir operator and, at leading order, can be related as
\be
\chi (p)^F \approx \left[ \frac{(c_F)^{\gamma_{F}}}{(c_G)^{\gamma_{G}}} \right] \chi (p)^G \,\, ,
\label{relate}
\ee
where $\chi (p)^i$, with $i=F,G$, indicates the $0^+$ scalar wave function formed by TC
fermions in the fundamental ($i=F$) or adjoint ($i=G$) representation.  Eq.(19) relates
the Bethe-Salpeter wave function for fermions in different representations of the
same gauge group. The Bethe-Salpeter equation (BSE) in the ladder approximation for
the scalar(or pseudo-scalar) channel is formally identical to the fermionic Schwinger-Dyson
equation (DSE). In these equations the interaction strength is proportional to
the Casimir operator ($c$), which is the only factor that is different for the fundamental and adjoint
representations in these equations. The factor $c$ will appear in front of the integral equation, and,
considering our ansatz in Eq.(9), it also appear in the exponent $\gamma (\alpha)$. For both representations
we will perform an integration (of the SDE or BSE) in the limit $\alpha \rightarrow 0$ and the result is less dependent
numerically on the factor $c$ in the exponent than it depends on this factor that is in
front of the integral, i.e. as a rough approximation we can assume that for the same gauge group the different
wave functions scale with $c$, in such a way that we can approximately obtain the fundamental wave function from
the adjoint one just changing the $c$ factors.

We now make the strong assumption that the magnitude of the condensates (or bound states) generated in the Minimal ($mi$) and Ultraminimal ($um$) models at one scale $\Lambda_{TC}$ and with fermions in the adjoint representation are the same, as long as
the gauge group is the same (no matter we have Dirac or Weyl fermions), implying that we can expect the following relation for the
scalar wave functions: 
\be
\chi (p)^G_{um} \approx \chi (p)_{mi} \,\, .
\label{relate1}
\ee
With this and Eq.(\ref{relate}) we may write
\be
\chi (p)^F_{um} \approx \left[ \frac{(c_F)^{\gamma_{F}}}{(c_G)^{\gamma_{G}}} \right]_{um} \chi (p)_{mi} \,\, .
\label{relate2}
\ee

The Ultraminimal model has two composite scalar bosons whose masses, in principle, would be given
by Eq.(\ref{eqf}), one formed by TC fermions in the fundamental representation and the other
formed by technifermions in the adjoint representation. 
We expect that the two scalar masses are not far apart, but following
our previous reasoning we will not compute them directly, but just say that the masses generated in the Ultraminimal model can also be related to the one of the Minimal model. Therefore, after  we consider  Eq.(\ref{relate1}) and  the normalization condition of the BSE we can estimate that the mass generated for the composite boson formed by TC fermions in the adjoint representation in the Ultraminimal model, which we denote by $M^{um}_{{H}_{1}}$, is approximately the mass generated for the composite boson in the Minimal model 
\be 
M^{um}_{{H}_{1}} \approx M^{mi}_{H} 
\ee 
\noindent where $M^{mi}_{H}$ is determined from Eq.(\ref{eqf}). The mass of the composite boson formed by TC fermions in the fundamental representation can be obtained in an similar way. Eq.(\ref{relate2}) allow us to obtain
\be 
M^{um}_{{H}_{2}} \approx  M^{mi}_{H}\left[ \frac{(c_F)^{\gamma_{F}}}{(c_G)^{\gamma_{G}}} \right]_{um} \,\, , 
\ee 
\noindent and in this case we indicate by $M^{um}_{{H}_{2}}$ the mass of the $H_2$ boson, which is the mass obtained by the lightest  composite boson, which is the one that couples to the particles of the Standard Model. Note that a factor $\left[\frac{(c_F)^{4\gamma_F}}{(c_G)^{4\gamma_G}}\right]_{{}_{um}}$ should also be introduced in the
calculation of $\lambda_{3{H}_2}$. The origin of the 4th power in this factor can be understood looking at the Appendix B of \cite{doff1}, where we have shown that $\lambda_{3H}  \propto  \Sigma^4$ (see Eq.(B4) of that reference), resulting from the couplings to the Bethe-Salpeter
wave function in the limit that the internal moment $q \rightarrow 0$, giving $3H$ $\chi$ factors, remembering that $\chi \propto \Sigma$, and an extra factor comes from the fermion mass that runs in the loop of Fig.(1) (also proportional to $\Sigma$), what leads to the scale factor
$(c_F)^{{}_{4\gamma_F}}/(c_G)^{{}_{4\gamma_G}}$ that we discussed above.

Therefore, based on the fact that the scalar wave functions are proportional to the Casimir operators, and that fermions in the same gauge group and representation have similar condensates, we were able to relate the Minimal model  and the Ultraminimal model and obtain information on their scalar masses.
\begin{table*}[t]
\caption{Scalar composite masses, trilinear couplings and decay widths into electroweak bosons of the Minimal and Ultraminimal TC models. In the Ultraminimal  model we show only the mass of the lightest Higgs $({H}_2)$. For  comparison we also include the same values for an ordinary $SU(2)$ walking theory with $8$ Dirac fermions in the fundamental representation. The $H_1$ boson mass is approximately the same one of the Minimal model. }
\vspace*{0.2cm}
\label{table1}
\newcommand{\m}{\hphantom{$-$}}
\newcommand{\cc}[1]{\multicolumn{1}{c}{#1}} 
\renewcommand{\tabcolsep}{2pc} 
\renewcommand{\arraystretch}{1.2} 
\begin{tabular}{@{}llll}
\hline 
 TC model & $M_H \,\,$(GeV) & $\lambda_{3H} \,\,$(GeV) & $\Gamma_{H \rightarrow WW}\,\,$(GeV) \\ 
 \hline
Walking & $142$ & $19$  &  $-$ \\
Minimal&  $414$ &  $16$ &  $109$ \\  
Ultraminimal $(H_2)$ & $250$ & $2.4$ & $17$ \\
\hline
\end{tabular}\\[2pt]
\end{table*}

Equation (\ref{relate2}) also leads to a simple determination of the decay width of $H_2$ into electroweak
gauge bosons in the case
of the Ultraminimal TC model. The decay width of the scalar boson into electroweak gauge bosons comes
from a loop coupling the composite scalar boson to two gauge bosons, but more precisely to the two
scalar wave functions of the (Goldstone) bosons absorbed by the electroweak gauge bosons. If the scalar wave
functions are related through Eq.(\ref{relate2}) we expect that the loop calculation only changes due
to the different wave functions factors that enter in the loop calculation. The decay width of the light
boson in the Ultraminimal model 
will be reduced by a factor proportional to a ratio of Casimir factors, and will be given by
\be
\Gamma_{H\rightarrow WW}^{um}\approx \Gamma_{H\rightarrow WW}^{mi}
\left[ \frac{(c_F)^{4\gamma_{F}}}{(c_G)^{4\gamma_{G}}} \right]_{um} (1-{O}(bg^2)...) \,\, .
\label{larg2}
\ee
\noindent To obtain the  decay width  for the Ultraminimal model  we consider  $c_F = 3/4$, $c_G = 2$  with  $\gamma_{i} = \frac{3c_i}{16\pi^2 b_{um}} $, for $i = F, G$. The scaling factor in Eq.(\ref{larg2}) appears because  $\Gamma_{HWW} \propto g^2_{{}_{HWW}}$  and
$g^2_{{}_{HWW}} \propto \Sigma^4$ (see Eq.(6) of the first paper in Ref.\cite{soni}). The decay width will be decreased considerably and only the composite boson formed by non-singlet technifermions under the electroweak group contribute to the decay.  The result of 
this calculation is in Table I ($\approx 17$GeV). If the calculation of the $H_2$ decay widht were performed with Eq.(\ref{larg}) the result would be slightly different ($\approx 24$GeV), indicating that the hypothesis  about the scalar wave functions and mass relations are quite reasonable.

Our results for the Higgs masses are in rough agreement with previous estimates. For instance, Ref.\cite{tuominen} contains an
extensive discussion about the scalar masses. Different arguments about the light scalar masses in these type of models can be 
found in the Section IV of Ref.\cite{dietrich}, and in the Appendix E of the second reference in \cite{tc}. Finally, a light Higgs can help 
to unitarize pi-pi scattering in these models (even though the Higgs contribution is not necessary)\cite{foadi}. Our results for the scalar bosons masses, trilinear couplings and decay widths in the Minimal and Ultraminimal TC models are displayed in Table I. For comparison we also add the mass and coupling of an ordinary walking $SU(2)_{TC}$ with $8$ Dirac fermions in the fundamental representation.  The trilinear self-coupling turn out to be much smaller than the one that we could expect when comparing with a fundamental Standard Model
scalar boson (where the coupling is $\lambda = 3 M_H^2/v$).
 
\section{Conclusions}

To conclude, we presented  a discussion about the  scalar composite masses, couplings and decay widths in the case of the Minimal and Ultraminimal TC models. 	To determine the mass generated for the Higgs boson in these models  we consider the BSE approach  developed in Ref.\cite{doff2} so that  we complement the results obtained in that work.  We also estimated the trilinear scalar self-coupling  for these models following
the calculation of Ref.\cite{doff1}. Our results are shown in Table I. 
These light scalar bosons can be produced at the LHC through the gluon-gluon fusion mechanism. Although the
isolated scalar production may be copious due to the small scalar mass, the possibility to observe pair production,
which has a substantial contribution from the trilinear coupling \cite{eboli} is not so favored due to the small values
of this coupling when compared to the one of an elementary scalar boson.

The composite scalars, being light, can also be produced in  association with a W/Z gauge boson and these channels can be both enhanced and feature more distinct final state distributions in walking technicolor models such as the Minimal and Ultraminimal, as compared to the Standard Model \cite{Mads2}. The Ultraminimal model also includes decays into un-eaten Goldstone bosons leading to very interesting invisible decays \cite{Mads1}. Note that our results were obtained without considering the mixing of the scalars, which can, in principle, be computed with the help
of an effective potential for composite operators, and this may still be complicated by the mixing with
scalars formed by technigluons, which is a problem far from being solved even in the QCD case.

The Minimal and Ultraminimal scalar masses are not so different. This is not surprising because both models are based
in the $SU(2)$ gauge group and most of the TC chiral symmetry breaking is triggered by adjoint technifermions,
and the Ultraminimal model contains an extra contribution coming from TC fermions in the fundamental
representation.  The Higgs boson formed by  TC fermions in the fundamental representation is light  and is
the one that couples to the electroweak gauge bosons.
The decay width into electroweak bosons  in the Ultraminimal model is quite different of the width obtained from the Minimal model,
because in the  Ultraminimal model the adjoint fermions are singlets under the electroweak coupling and only the
scalar bound state that is formed with fundamental fermions contributes to this decay.

\section*{Acknowledgments}

We thank F. Sannino for calling our attention to the models discussed in this work and P. S. Rodrigues da Silva for useful discussions. This research was partially supported by the Conselho Nacional de Desenvolvimento Cient\'{\i}fico e Tecnol\'ogico (CNPq).

\end{document}